\documentclass[11pt]{article} 

\usepackage{amsfonts}
\usepackage{amsmath}
\usepackage{amssymb}
\usepackage{amsthm}
\usepackage{caption}
\usepackage{color}
    \definecolor{darkgreen}{rgb}{0,0.5,0}
    \definecolor{darkblue}{rgb}{0,0,0.6}
    \definecolor{purple}{rgb}{0.4,.2,0.7}
\usepackage[margin = 2.5cm]{geometry}
\usepackage{graphicx}
\usepackage[hyperfootnotes = false, colorlinks = true, linkcolor = darkblue, citecolor = blue,urlcolor  = darkblue]{hyperref}
\usepackage{subcaption}
\usepackage{csquotes}

\usepackage{braket}
\usepackage{tikz}
\allowdisplaybreaks[4]
\newcommand{\bea}{\begin{eqnarray}}
\newcommand{\eea}{\end{eqnarray}}
\def\la{\label}

\begin{document}
\begin{titlepage}
	\renewcommand{\thefootnote}{\fnsymbol{footnote}}
\thispagestyle{empty}
\begin{center}
    ~\vspace{5mm}
    
    {\Large \bf {Joule-Thomson effect of AdS black holes in conformal gravity}}

    \vspace{0.5in}
    
    {\bf Yang Guo${}^{}$\footnote{guoy@mail.nankai.edu.cn}, Hao Xie${}^{}$\footnote{xieh@mail.nankai.edu.cn}, and Yan-Gang Miao${}^{}$\footnote{Corresponding author: miaoyg@nankai.edu.cn}}

   \vspace{0.5in}

   {\em School of Physics, Nankai University, Tianjin 300071, China}
     
    \vspace{0.5in}

\end{center}

\vspace{0.5in}

\begin{abstract}
  We investigate the Joule-Thomson effect of AdS black holes in conformal gravity. 
  We derive the Joule-Thomson coefficient in terms of  thermodynamic relations and then make an alternative derivation via a direct way. We analyze the Joule-Thomson coefficient and find that the Joule-Thomson coefficients obtained  from two different approaches are equal. Moreover, we present a novel isenthalpic process in which the inversion temperature is minimal and it separates the corresponding heating-cooling phase. We analyze the inversion temperature and its corresponding inversion curve that separates the regions for the JT effect to be allowable and forbidden, where such an effect can only be observed in the allowable region.
  We also discuss the effects of two important parameters on the inversion curves.
\end{abstract}

\vspace{1in}


\setcounter{tocdepth}{3}


\end{titlepage}
\tableofcontents
\renewcommand{\thefootnote}{\arabic{footnote}}
\setcounter{footnote}{0}
\section{Introduction} 
     
 The black hole thermodynamics has always been a topic of great concerns and challenges over the past five decades. A  black hole is no longer a mechanical system described formally by an analogy with thermodynamics but actually a thermodynamic system with temperature and entropy since the important discoveries  were made, such as the Hawking area theorem~\cite{Hawking:1971vc}, the Bekenstein entropy~\cite{Bekenstein:1973ur}, and the Hawking radiation~\cite{Hawking:1975vcx}.  In an extended phase space, the cosmological constant is viewed~\cite{Caldarelli:1999xj,Kastor:2009wy,Dolan:2010ha} as a thermodynamic variable and identified as the thermodynamic pressure. This
 has led to a wide study in black hole thermodynamics, where many phenomenological aspects of thermodynamics are introduced into black hole physics.

 A number of phenomenological studies show that black holes behave in many ways quite analogous to common thermodynamic phenomena, such as the liquid/gas phase transition~\cite{Poshteh:2013pba,Cai:2013qga}, the van der Waals fluid behavior~\cite{Hennigar:2015wxa,Zeng:2015wtt,Guo:2021wcf}, the triple point~\cite{Wei:2014hba,Frassino:2014pha,Altamirano:2013uqa}, and the heat engines~\cite{Johnson:2014yja,Belhaj:2015hha,Johnson:2016pfa,Guo:2022cdj}, {\em etc}. Furthermore, we can consider the adiabatic expansion in the field of black hole thermodynamics in which the cosmological constant is treated as a thermodynamic pressure. In the common thermodynamics, the adiabatic expansion usually refers to the well-known Joule-Thomson (JT) effect, which describes the change in temperature of a gas or liquid when the gas or liquid  passes through a throttle, like a porous plug or a small valve. This process is known as throttling or JT expansion, in which the enthalpy remains unchanged. We note that the JT expansion is analyzed~\cite{Okcu:2016tgt} for the first time in charged AdS black holes, where the cooling and heating regions are clearly shown. Moreover, the similarity and difference between charged black holes and van der Waals fluids are also discussed~\cite{Okcu:2016tgt} in the aspect of adiabatic expansion. Recently, the JT expansion of black holes has been extensively studied~\cite{Kuang:2018goo,Xing:2021gpn,Cao:2022hmd,Barrientos:2022uit,Assrary:2022uiu,Nam:2019zyk,Bi:2020vcg,Zhao:2018kpz} in different gravity theories. In the present work, we generalize the JT effect to AdS black holes in conformal gravity with the motivations described below.

 When a black hole is dealt with as a thermodynamic system, its further investigations provide insights into  the fundamental relationship among gravity, thermodynamics, and quantum physics. We analyze such a JT expansion that occurs in the AdS black holes of conformal gravity~\cite{Riegert:1984zz,Klemm:1998kf}, where the characteristic of this model relies on the mass which is introduced~\cite{Lu:2012xu,Li:2012gh} by a time-like Killing vector and its corresponding Noether charge.
 Here the characteristic means that the mass parameter does not appear explicitly in the metric function of the model but as a specific combination~\cite{Xu:2014kwa,Xu:2018vxf} of the parameters appeared in this metric function. This feature distinguishes the conformal gravity from the other theories of  gravity in which the  mass is simple and well-formed in metrics.  The JT effect we consider includes the heating-cooling effect and inversion temperature. On one hand, we want to verify whether the mass and the other thermodynamic quantities of AdS black holes in conformal gravity are well-defined during the JT process. On the other hand, we want to reveal some effects of conformal  gravity on the JT expansion. More importantly, the novel phase behavior of AdS black holes in conformal gravity can extend the burgeoning field of black hole thermodynamics.

Our paper is organized as follows. In Sec.~\ref{sec:thermodynamics} we make a brief review for AdS black holes in conformal gravity and the relevant thermodynamics.  In Sec.~\ref{subsec:JTc} we derive the analytic expression of JT coefficients in two different ways, where one is based on thermodynamic relations and the other is a direct derivation.  In Sec.~\ref{subsec:isen} we present a novel isenthalpic process and discuss the effects of two important parameters on inversion curves.
Finally, we give our conclusions and discussions in Sec.~\ref{sec:con}

\section{AdS black holes in conformal gravity and relevant thermodynamics}
\label{sec:thermodynamics}
A conformal gravity, constructed from the quadratic Weyl term, may be an alternative theory of gravitation, and its Lagrangian reads~\cite{Riegert:1984zz,Klemm:1998kf}
\begin{eqnarray}
\mathcal{L} = \frac{1 }{2} \alpha C^{\mu\nu\rho\sigma} C_{\mu\nu\rho\sigma},\label{Lconformal}
\end{eqnarray}
where $C_{\mu\nu\rho\sigma}$ is the Weyl tensor and the sign choice of coupling constant $\alpha$ is critical~\cite{Lu:2011zk} though $\alpha$ is independent of the equations of motion. The conformal gravity admits~\cite{Lu:2012xu} the following general static black hole solutions,
\begin{eqnarray}
 	ds^2=-f(r)dt^2+\frac{dr^2}{f(r)}+r^2d\Omega_{2,\epsilon}^2,\nonumber
\end{eqnarray}
\bea
 	f(r)=c_1 r+c_0+\frac{d}{r}-\frac{\Lambda  r^2}{3},\label{shapefunc}
\eea
where $d\Omega_{2,\epsilon}^2 $ is the line element of a unit hyperbolic two-space $H^2$, or a torus $T^2$, or a sphere $S^2$, which corresponds to $\epsilon=-1,0,1$, respectively, and the parameters $c_0$, $c_1$ and $d$ are constants constrained by
\bea
	3c_1d+\epsilon^2=c_0^2.
\eea
In the following discussions we take $\epsilon=1$, i.e., the topology of two-spheres is chosen. Moreover, we regard $c_0$ and $d$ as independent parameters whose physical meanings in conformal gravity have been given in Ref.~\cite{Lu:2012xu}.  

The mass of AdS black holes in conformal gravity can be calculated~\cite{Lu:2012xu,Li:2012gh} by defining the conserved charge associated with the time-like Killing vector, which should be identified with the enthalpy,
\bea
 	H=\frac{\alpha}{24
 		\pi r_+}\left[ \frac{ \left(-c_0+\epsilon+2 \Lambda  r_+^2 \right)d}{r_+}+\frac{\left(c_0-\epsilon \right) \left(-3 c_0+\Lambda  r_+^2\right)}{3 }\right],\label{eq:H}
\eea
where $r_+$ is the largest root of $f(r)=0$, and the Hawking temperature can thus be obtained,
\bea
 	T=\frac{-3 c_0 r_+-6 d+8 \pi  P r_+^3}{12 \pi  r_+^2},\label{eq:T}
\eea
where the pressure is introduced,
\bea
P=-\frac{\Lambda}{8\pi},\label{eq:P} 
\eea
in the extended phase space. Therefore, we can write the first law of black hole thermodynamics,
\begin{eqnarray}
	dH=TdS +\Psi d\Xi+VdP,\label{eq:dH}
\end{eqnarray}
where $\Xi=c_1$ and its conjugate $\Psi$ is viewed as a massive spin-2 hair. According to the explicit expressions of enthalpy $H$ and temperature $T$ given in Eqs.~\eqref{eq:H} and \eqref{eq:T}, we can calculate the heat capacity at constant pressure by employing the standard thermodynamic method,
\bea
    	C_P=\left( \frac{\partial H}{\partial T}\right)_{P,\Xi}= \frac{\alpha\left(c_0 -\epsilon\right) \left(3 c_0 r_+ +6 d-8 \pi  P r_+^3\right)}{6 \left(3 c_0 r_+ +12 d+8 \pi  P r_+^3\right)}.\label{cp}
\eea
 \section{Joule-Thomson expansion}
In the common thermodynamics the well-known Joule-Thomson (Joule-Kelvin) effect describes the change of temperature of a gas or liquid when the gas or liquid passes through a throttle, such as a porous plug or a small valve, in a thermally insulated environment. This procedure is called the throttling process or Joule-Thomson expansion in which the enthalpy $H$ remains unchanged. In this section, we generalize the Joule-Thomson effect to the thermodynamics of AdS black holes in conformal gravity  and investigate the peculiar behavior of such a phenomenon in the model governed by Eq.~(\ref{Lconformal}). As in the common thermodynamics, a physical effect can usually be depicted by a partial derivative, so one defines
\bea
	\mu=\left(\frac{\partial T}{\partial P} \right)_H 
\eea
to represent the rate of change of temperature $T$ with respect to pressure $P$ at constant enthalpy $H$, where $\mu$ is called the Joule-Thomson coefficient.
\subsection{Two approaches to derive Joule-Thomson coefficients}
\label{subsec:JTc}
\subsubsection{A derivation based on the relations between the Joule-Thomson coefficient and thermodynamic quantities}

The entropy $S$ is a state function and  can be viewed as the function of  the pressure $P$, Hawking temperature $T$ and  massive spin-2 hair $\Xi$. So its total differential formula takes the form,
\bea
	dS=\left(\frac{\partial S}{\partial P} \right)_{T,\Xi}dP +\left(\frac{\partial S}{\partial T} \right)_{P,\Xi}dT  +\left(\frac{\partial S}{\partial \Xi} \right)_{T,P}d\Xi.\la{eq:dS}
\eea
Substituting this expression into Eq.~\eqref{eq:dH} and using the condition $d\Xi=0$, we find
\bea
	dH=T\left( \frac{\partial S}{\partial T}\right)_{P,\Xi}dT+\left[\left(T \frac{\partial S}{\partial P}\right)_{T,\Xi} +V \right] dP,
\eea
which can be rearranged to give
\bea
C_P=\left( \frac{\partial H}{\partial T}\right)_{P,\Xi}=T\left( \frac{\partial S}{\partial T}\right)_{P,\Xi},\label{eq:cp}
\eea
\bea
	\left( \frac{\partial H}{\partial P}\right)_{T,\Xi}=T\left( \frac{\partial S}{\partial P}\right)_{T,\Xi}+V.\la{eq:dhdp}
\eea
As a result, we obtain  the heat capacity at constant pressure from Eq.~\eqref{eq:cp}. Moreover, using Maxwell's relation,
\bea
		\left( \frac{\partial S}{\partial P}\right)_{T,\Xi}=-\left( \frac{\partial V}{\partial T}\right)_{P,\Xi},
\eea
we rewrite Eq.~\eqref{eq:dhdp} as
\bea
	\left( \frac{\partial H}{\partial P}\right)_{T,\Xi}=-T\left( \frac{\partial V}{\partial T}\right)_{P,\Xi}+V.\label{eq:dhdp1}
\eea

We note that the JT coefficient depends on the three variables ($T$, $P$, $H$) and the enthalpy as a state function can be expressed as $H=H(T,P)$. By applying the cyclic rule we obtain the following useful formula,
\bea
	\left( \frac{\partial T}{\partial P}\right)_{H}\left( \frac{\partial P}{\partial H}\right)_{T,\Xi}\left( \frac{\partial H}{\partial T}\right)_{P,\Xi}=-1.\label{eq:rule}
\eea
Finally, we work out the following expression of the JT coefficient by using Eqs.~\eqref{eq:cp}, \eqref{eq:dhdp1}, and \eqref{eq:rule},
\bea
	\mu=\left(\frac{\partial T}{\partial P} \right)_H =\frac{1}{C_P}\left[T \left( \frac{\partial V}{\partial T}\right)_{P,\Xi} -V\right], \label{eq:mut}
\eea
which shows a clear relationship between the JT coefficient and thermodynamic quantities. Taking the thermodynamic quantities into account in Eq.~\eqref{eq:mut}, we can immediately obtain the exact JT coefficient once the heat capacity $C_P$ and the other thermodynamic quantities are identified. Thus, Eq.~\eqref{eq:mut} is  a very helpful formula to express the JT coefficient in terms of the other thermodynamic quantities that can  conveniently be measured.

\subsubsection{A direct derivation}
In this subsection, we compute the JT coefficient alternatively by using a direct way. We first write the pressure $P$ and Hawking temperature $T$ as the function of  enthalpy $H$ and event horizon $r_+$, respectively,
\bea
	P(H,r_+)=\frac{3 \left(\alpha  c_0 d  +\alpha  c_0^2 r_+  -\alpha  c_0 r_+  \epsilon -\alpha  d   \epsilon +24 \pi  H r_+^2\right)}{8 \pi  \alpha  r_+^2   \left(-c_0 r_+-6 d+r_+ \epsilon \right)},\label{eq:phr}
\eea

\bea
T(H,r_+)=\frac{\alpha  c_0 r_+  \left(2 c_0 r_++9 d-2 r_+ \epsilon \right)+3 \alpha  d   \left(4 d-r_+ \epsilon \right)+24 \pi  H r_+^3}{4 \pi  \alpha  r_+^2 \left(-c_0 r_+-6 d+r_+ \epsilon \right)}.\label{eq:thr}
\eea
With the aid of the above two equations, we then give the JT coefficient directly,
\begin{eqnarray}
\mu=\left( \frac{\partial T}{\partial P}\right)_H= \frac{-4 c_0 r_+ \left(c_0 r_+ +6 d-r_+ \epsilon \right)-4 d \left(12 d+8 \pi  P r_+^3-3 r_+ \epsilon \right)}{\left(\epsilon -c_0\right) \left(3 c_0 r_++6 d-8 \pi  P r_+^3\right)}\label{mu}.
\end{eqnarray}

We can see that the JT coefficient is independent of coupling constant $\alpha$. It is also easy to check that the JT coefficients depicted by Eq.~\eqref{eq:mut} and Eq.~\eqref{mu} are equivalent to each other, where what we need to do is to substitute the required thermodynamic quantities, such as the heat capacity and temperature, into Eq.~\eqref{eq:mut}. 

 \subsection{A novel isenthalpic process and inversion curve on $T-P$ plane}
 \label{subsec:isen}
\begin{figure}[h]
\begin{center}
\includegraphics{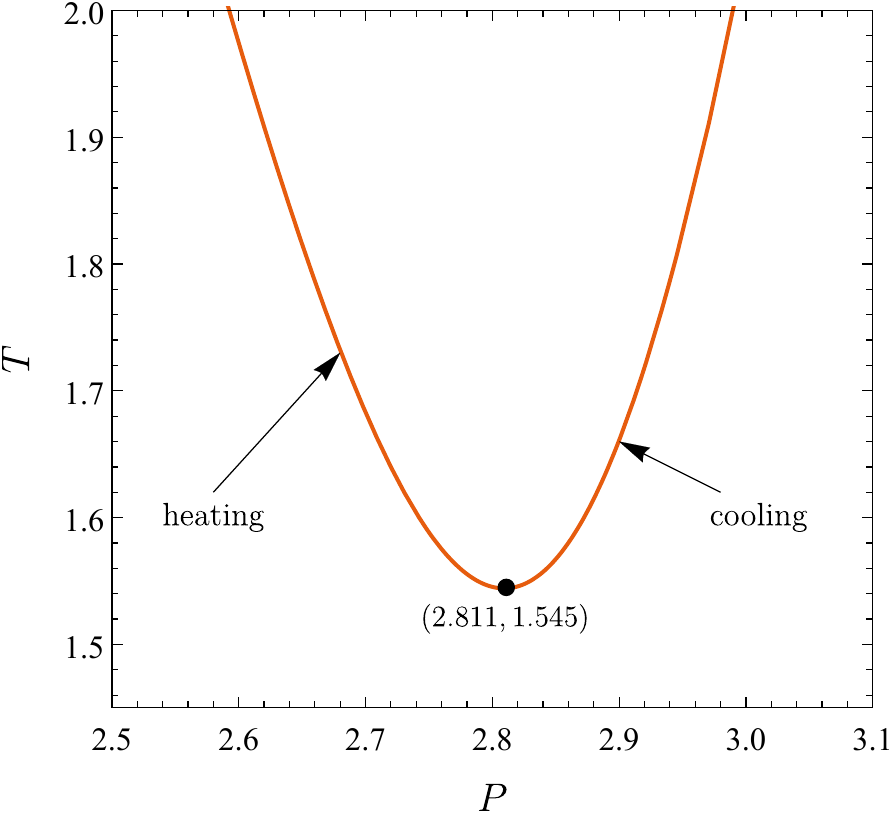}
\end{center}
\caption{The isenthalpic process for AdS black holes in conformal gravity with $d=-1$, $c_0=1/10$, $\epsilon=1$, and $H=2$. The black point on the curve corresponds to the inversion point, $(P_{\rm inv}, T_{\rm inv})=(2.811, 1.545)$, in the isenthalpic process with $\mu=0$.}
\label{fig:isenthalpy}
\end{figure}

 The JT expansion occurs in a thermally insulated environment. During this expansion, the enthalpy $H$ remains unchanged. Utilizing Eqs.~\eqref{eq:phr} and \eqref{eq:thr}, we present an isenthalpic process in Fig.~\ref{fig:isenthalpy},  which allows us to give the corresponding heating-cooling phase. The JT coefficient $\mu$ corresponds to the slope on the isenthalpic curve. On the right half of the curve ($\mu>0$), the temperature drops when the model we consider goes through a throttling process. Therefore,  the right half part is called cooling region because the temperature is decreasing with a decrease of the pressure. In contrast, on the left half of the curve ($\mu<0$), the temperature rises after throttling and this half part is called heating region. The inversion point, ($P_{\rm inv}$, $T_{\rm inv}$) at $\mu=0$, separates the cooling and heating regions where $T_{\rm inv}$  is the inversion temperature given by 
 \begin{eqnarray}
 	T_{\rm inv}=V\left(\frac{\partial T}{\partial V} \right)_{P}.\label{eq:ti} 
 \end{eqnarray}

We point out that $T_{\rm inv}$ is the minimum temperature that the AdS black hole of conformal gravity can reach in the JT process, which is opposite to the phenomenon observed in the other black hole models~\cite{Lan:2018nnp,Mo:2018rgq,Cisterna:2018jqg,Li:2019jcd,Lan:2019kak,Guo:2020qxy} and the van der Waals fluid~\cite{Okcu:2016tgt,Okcu:2017qgo} where the corresponding temperature is the maximum during the JT expansion. We reveal the peculiar behavior for the AdS black hole of conformal gravity in the aspect of inversion temperature.   

\begin{figure}[h]
\begin{center}
\includegraphics{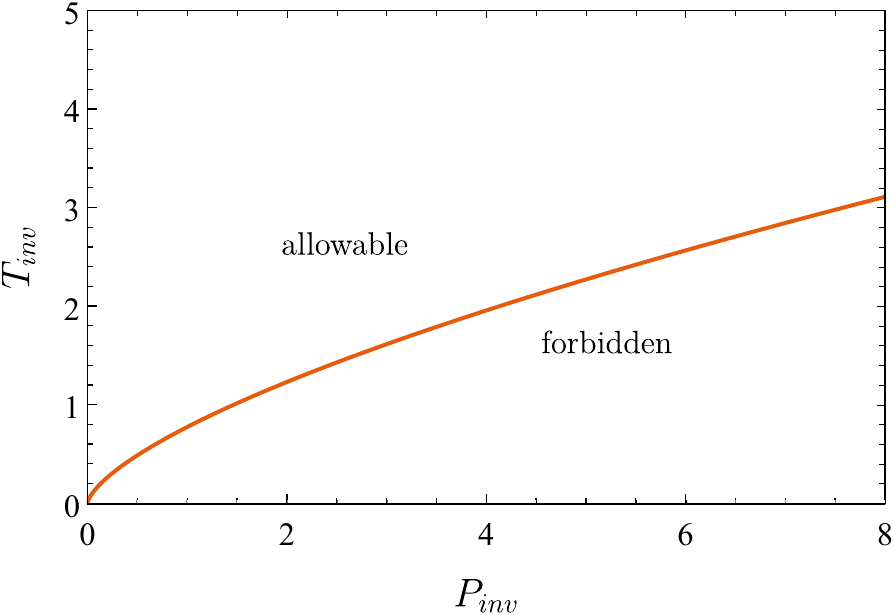}
\end{center}
\caption{The inversion curve for AdS black holes in conformal gravity with  $d=-1$, $\epsilon=1$, and $c_0=1/10$. The JT expansion occurs in the allowable region.}
\label{fig:af}
\end{figure}

\begin{figure}[h]
\begin{center}
\includegraphics{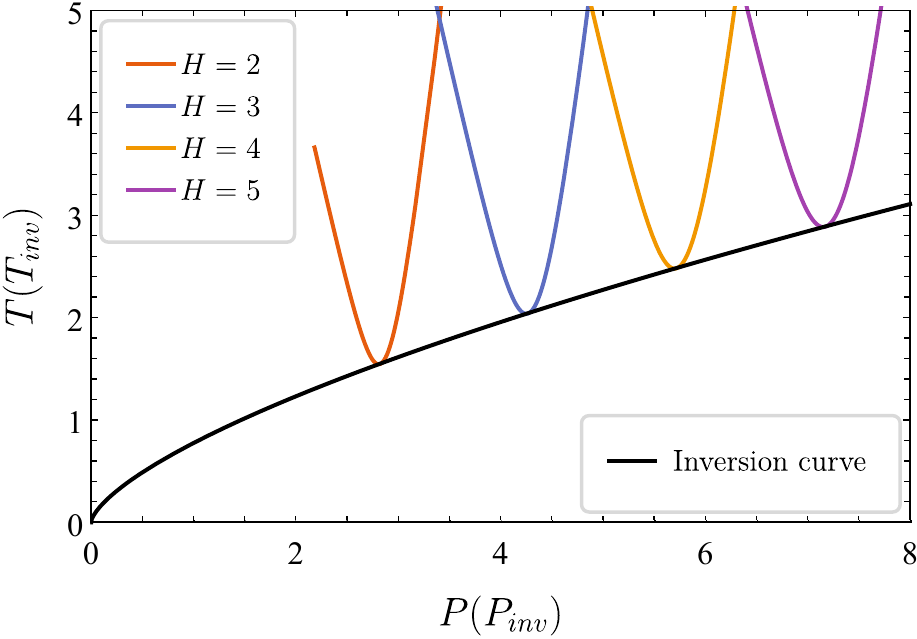}
\end{center}
\caption{The isenthalpic and inversion curves for AdS black holes in conformal gravity with  $d=-1$, $\epsilon=1$, and $c_0=1/10$.}
\label{fig:ttpp}
\end{figure}

Next, we analyze the other peculiar behavior in the aspect of inversion curves. An inversion curve with the setting, $\mu=0$, can be obtained from Eq.~\eqref{eq:ti}, which is shown in Fig.~\ref{fig:af} (red curve) and Fig.~\ref{fig:ttpp} (black curve). The isenthalpic curves are shown as colored curves for various values of enthalpy, see Fig.~\ref{fig:ttpp}, where the inversion point, $(P_{\rm inv}, T_{\rm inv})$, rises monotonically as the enthalpy increases. The inversion curve is depicted by the locations of isenthalpic curves that have a vanishing slope.   Here we can observe a significant feature that the JT expansion of AdS black holes in conformal gravity experiences a cooling process at first, and then a heating process if a large enough initial pressure is given. Additionally, the inversion curve no longer crosses these isenthalpic curves but makes a tangency to them, which presents a different behavior from that of the charged AdS black holes~\cite{Okcu:2016tgt}. Our observation suggests that each point on the inversion curve marks the boundary, above which the JT expansion can occur but below which the JT expansion cannot occur. Thus, our inversion curve denotes the dividing curve between the allowable and forbidden regions for the JT expansion as shown in Fig.~\ref{fig:af}, while that inversion curve in the black hole models~\cite{MahdavianYekta:2019dwf,Liang:2021xny,Kruglov:2022mde,Rajani:2020mdw,Ghaffarnejad:2018tpr,Feng:2020swq} mean the dividing curve between the cooling and heating regions. As a result, the JT expansion  will only occur in the parameter space above the inversion curve in our model.

\begin{figure}[h]
\centering
\begin{subfigure}[b]{0.48\textwidth}
\centering
\includegraphics[width=\textwidth]{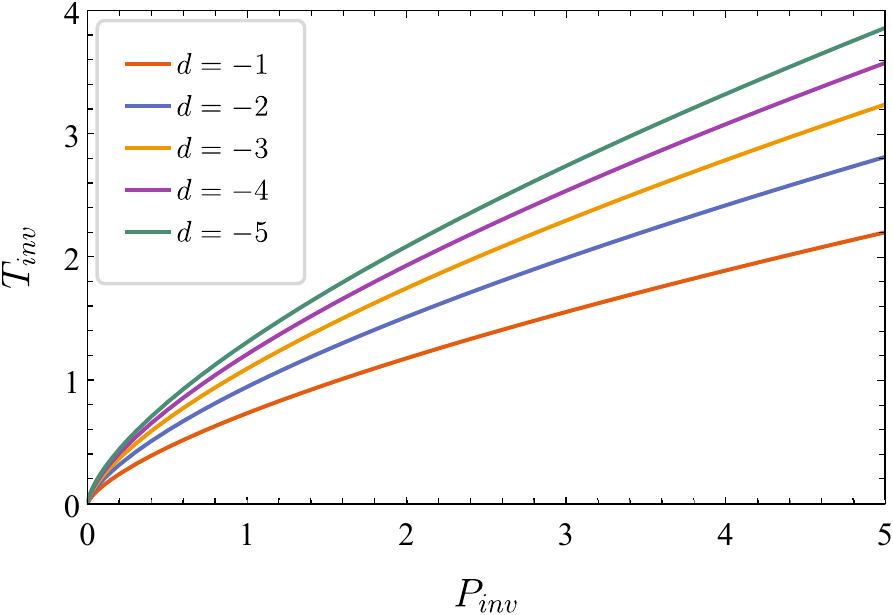}
\end{subfigure}
\begin{subfigure}[b]{0.512\textwidth}
\centering
\includegraphics[width=\textwidth]{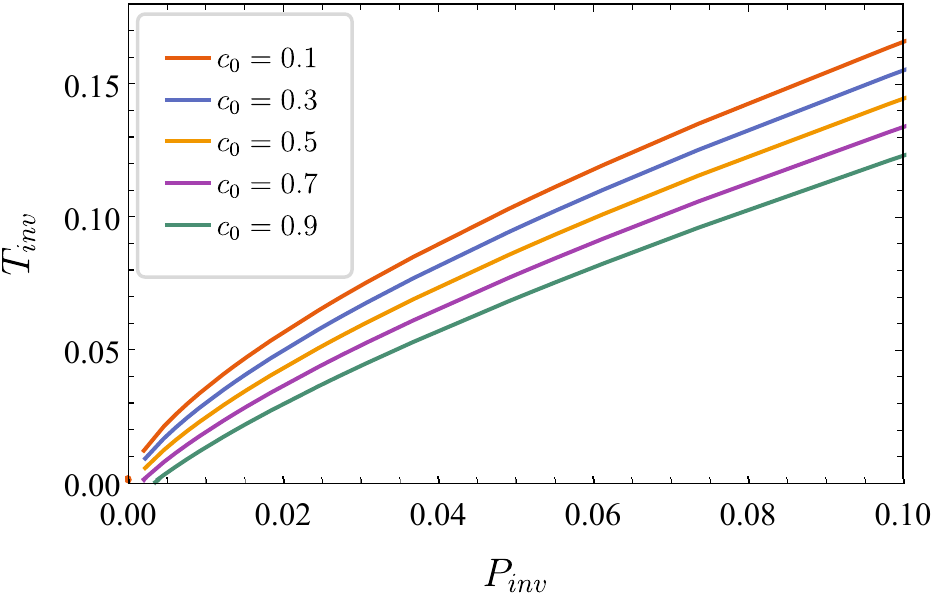}
\end{subfigure}
\caption{The dependence of the inversion curves on the parameters $d$ and $c_0$. The left panel gives the inversion curves for $d=-1, -2, -3, -4, -5$ with a fixed $c_0=1/2$. The right panel presents the inversion curves for $c_0=0.1, 0.3, 0.5, 0.7, 0.9$ with a fixed $d=-1$.}
\label{fig:dc0}
\end{figure}

The relationship between the inversion curve and parameter $d$ for a fixed parameter $c_0=1/2$ is depicted by the left diagram of Fig.~\ref{fig:dc0}. The inversion temperature increases with a decrease of the parameter $d$, which indicates that the smaller the parameter $d$ is, the earlier the transition between the cooling and heating processes is achieved in the JT process of conformal gravity. The right diagram of Fig.~\ref{fig:dc0} shows the relationship between the inversion curve and parameter $c_0$ with a fixed parameter $d=-1$. 
The inversion temperature drops with an increase of the parameter $c_0$, which shows that the larger the parameter $c_0$ is, the later the transition between the cooling and heating processes is realized in the JT process.   
   
\section{Conclusion and discussion} 
\label{sec:con}
In this work we investigate the JT expansion for AdS black holes in conformal gravity. We derive the  expression of JT coefficients in terms of thermodynamic relations at first, and then we make the derivation via a direct way. We show that the results from two different approaches are equal, which indicates that the enthalpy and first law of thermodynamics are well-defined for the AdS black holes in conformal gravity. 
Our results show a clear feature that the JT coefficient is independent of coupling constant $\alpha$ that appears in the Lagrangian of conformal gravity, indicating that the JT expansion we consider here is universal.

In particular, we find a novel isenthalpic process in which the minimum temperature of the JT expansion, rather than the maximum one reported by previous works~\cite{Okcu:2016tgt,Lan:2018nnp,Mo:2018rgq,Cisterna:2018jqg,Li:2019jcd,Okcu:2017qgo}, separates the corresponding heating-cooling phase. This process shows that the AdS black hole in conformal gravity goes through the cooling phase and then through the heating phase during the throttling process. Therefore,  our inversion curve separates the allowable and forbidden regions for the JT effect to be observed, but not the cooling and heating regions reported by previous works~\cite{MahdavianYekta:2019dwf,Liang:2021xny,Kruglov:2022mde}. These novel behaviors depend on the characteristics of conformal gravity, where one of them is that the mass is no longer an independent variable but a special combination of black hole parameters,\footnote{As a comparison, for instance, the mass is explicitly  included in the metric as a independent variable in Einstein's gravity.} see Eq.~\eqref{eq:H}. 
Moreover, our results show that  two important parameters $d$ and $c_0$ have similar effects on the inversion curves, {\em i.e.}, the inversion temperature will drop with an increase of $d$ and $c_0$, which indicates a delay in the transition from a cooling to heating phase.

\paragraph{Acknowledgments}
The authors would like to thank the anonymous referee for the helpful comments that improve
this work greatly.
This work was supported in part by the National Natural Science Foundation of China under Grant No. 12175108. 
      
\appendix 

\bibliographystyle{JHEP}
\bibliography{references}
 
\end{document}